\documentclass[prl,twocolumn,showpacs,superscriptaddress]{revtex4}
\usepackage{graphicx,psfrag,amsmath,amssymb,amsfonts,bbm}

\newcommand{\ele}[3]{\langle #1 \mid #2\mid #3\rangle}
\newcommand{\ket}[1]{|#1\rangle}

\newcommand{\ba}{\begin{eqnarray}}
\newcommand{\ea}{\end{eqnarray}}
\newcommand{\bary}{\begin{array}}
\newcommand{\ear}{\end{array}}

\begin{document}
\title{Interaction induced localization in a gas of pyramidal molecules} 
\author{Giovanni Jona-Lasinio}
\affiliation{Dipartimento di Fisica, Universit\`a di Roma ``La Sapienza'',
Piazzale Aldo Moro 2, Roma 00185, Italy}
\affiliation{Istituto Nazionale di Fisica Nucleare, Sezione di Roma 1, Roma 00185, Italy} 
\author{Carlo Presilla}
\affiliation{Dipartimento di Fisica, Universit\`a di Roma ``La Sapienza'',
Piazzale Aldo Moro 2, Roma 00185, Italy}
\affiliation{Istituto Nazionale di Fisica Nucleare, Sezione di Roma 1, Roma 00185, Italy} 
\affiliation{Istituto Nazionale per la Fisica della Materia, 
Unit\`a di Roma 1 and Center for Statistical Mechanics and Complexity, Roma 00185, Italy} 
\author{Cristina Toninelli}
\affiliation{Dipartimento di Fisica, Universit\`a di Roma ``La Sapienza'',
Piazzale Aldo Moro 2, Roma 00185, Italy}

\date{January 28, 2002}
\begin{abstract} 
We propose a model to describe a gas of pyramidal molecules interacting 
via dipole-dipole interactions. 
The interaction  modifies the tunneling properties between the classical 
equilibrium configurations of the single molecule and, 
for sufficiently high pressure, the molecules
become localized in these classical configurations. 
We explain quantitatively, 
without free parameters, the shift to zero-frequency of the 
inversion line observed upon increase of the 
pressure in a gas of ammonia or deuterated ammonia. 
For sufficiently high pressures, our model suggests the existence 
of a superselection 
rule for states of different chirality in substituted derivatives.  
\end{abstract}
\maketitle

The behavior of gases of pyramidal molecules, 
i.e. molecules of the kind $XY_3$ like ammonia $NH_3$,
has been the object of investigations since the early developments
of quantum mechanics \cite{Hund}.
However, the behavior of these systems is still debated and some 
experimental facts remain essentially unexplained \cite{Wightman}.

For the single pyramidal molecule, owing to the great differences 
of characteristic energies and times, 
some adiabatic approximations hold. 
In particular, the one-dimensional inversion motion 
of the nucleus $X$ across the plane containing the three nuclei $Y$
can be separated from the rotational and vibrational nuclear degrees
of freedom. 
The form of the effective potential for this motion is 
a double well which is symmetric with respect to the inversion plane 
\cite{Townes}.
Due to tunneling across the finite potential barrier, the eigenstates 
are delocalized in the two minima of the potential and, 
for energies below the barrier height, 
are grouped in doublets, i.e. couples of states with a relative splitting 
in energy small in comparison with the distance from the rest of the 
spectrum. 
For the pyramidal molecules under consideration,
the thermal energy $k_BT$ at room temperature is much smaller 
than the distance between the first and the second doublet so that  
the problem can be reduced to the study of a two-level system 
corresponding to the symmetric and anti-symmetric states of the 
first doublet. 
 
The existence of delocalized stationary states is clearly in 
disagreement with the usual chemical view which, 
relying upon classical theory, considers the molecules as objects 
with a well defined spatial structure. 
In particular, for the molecules under consideration 
the classical view predicts one of the two pyramidal configurations 
corresponding to the nucleus $X$ localized in one of the wells of the 
inversion potential. 

The quantum prediction of stationary delocalized states implies 
the presence of a line in the absorption spectrum, the so called 
inversion line, at a frequency $\bar\nu = \Delta E/h$, 
where $\Delta E$ is
the energy splitting of the first doublet. 
Experiments performed with $NH_3$ \cite{NH3}, $ND_3$ \cite{ND3} 
and $NT_3$ \cite{NT3} reveal the existence of this inversion line
in various rotational and vibrational bands.
The frequency $\bar\nu$ of the inversion line has been measured 
as a function of the gas pressure $P$ 
for $NH_3$ \cite{BL1,BL2} and $ND_3$ \cite{Birnbaum}.
Starting from the expected value $\Delta E/h$ at $P \simeq 0$,  
$\bar\nu(P)$ decreases by increasing $P$ and vanishes at a critical 
pressure $P_\mathrm{cr} \simeq 1.7$ atm for $NH_3$ 
and $P_\mathrm{cr} \simeq 0.1$ atm for $ND_3$. 
No quantitative theory has been proposed so far for this phenomenon.

As early as 1949, in a short qualitative paper \cite{Anderson}
Anderson made the hypothesis that dipole-dipole 
interaction may induce a localization of the molecular states.
In this way the important idea was introduced that inter-molecular 
interactions may be responsible for the observed phenomena.
By considering a system of two or three interacting molecules, 
Margenau was able to predict a decrease of the inversion frequency 
on reducing the distance between the molecules\cite{Margenau}.
A quantitative discussion of the collective effects induced by coupling 
a single molecule to the environment constituted by the other 
molecules of the gas was made in \cite{jonaclaverie}.
In this work it was shown that, due to the instability
of tunneling under weak perturbations,
the order of magnitude of the molecular dipole-dipole interaction 
may account for localized ground states.
This suggests that a kind of phase transition may be invoked 
to explain the behavior of $NH_3$ and $ND_3$ under variation 
of pressure.  

We have implemented this idea by constructing a 
simplified model of a gas of pyramidal molecules which exhibits the
desired properties and allows a direct comparison with experimental
data. 
Our model predicts, for sufficiently high inter-molecular interactions, 
the presence of two degenerate ground states corresponding to the  
different localizations of the molecules. 
This transition to localized states gives a reasonable 
explanation of the experimental results 
\cite{BL1, BL2, Birnbaum}. 
In particular, it describes quantitatively, without free parameters, 
the shift to zero-frequency of the inversion line of $NH_3$ and $ND_3$ 
on increasing the pressure.

We model the gas as a system of molecules nearly independent 
in the following sense:
each molecule is subjected to an external field representing 
the interaction with the rest of the gas to be determined 
self-consistently.
We then analyze the linear response of this model to an 
electromagnetic perturbation to obtain the low energy excitation 
spectrum and its dependence on the inter-molecular interaction. 
Finally, we compare our theoretical results with the available
experimental data.

We mimic the inversion degree of freedom of an isolated molecule with 
the Hamiltonian
\begin{equation}
h_0=-\frac{\Delta E}{2}\sigma^x,
\end{equation}
where $\sigma^x$ is the Pauli matrix in the standard representation
with delocalized tunneling eigenstates
\begin{equation}
\ket{1} = \frac{1}{\sqrt 2}
\left( \begin{array}{c} 1 \\ 1 \end {array} \right) 
\qquad
\ket{2} = \frac{1}{\sqrt 2} 
\left( \begin{array}{c} \phantom{-}1 \\ -1 \end {array} \right).
\label{12}
\end{equation}
Since the rotational degrees of freedom of the single pyramidal molecule 
are faster than the inversion ones, 
on the time scales of the inversion dynamics the molecules 
feel an effective attraction arising from the angle averaging of the 
dipole-dipole interaction at the temperature of the experiment 
\cite{Keesom}.
In the representation chosen for the Pauli matrices, 
the localizing effect of the dipole-dipole interaction between two
molecules $i$ and $j$ can be represented by an interaction term 
of the form $\sigma^z_i \sigma^z_j$, 
where $\sigma^z$ has localized eigenstates
\begin{equation}
\ket{L} = 
\left( \begin{array}{c} 1 \\ 0 \end {array} \right) 
\qquad
\ket{R} = 
\left( \begin{array}{c} 0 \\ 1 \end {array} \right).
\label{LR}
\end{equation} 
In a mean-field approximation we obtain the total Hamiltonian
\begin{equation}
h(\lambda)=-\frac{\Delta E}{2}\sigma^x-
G\sigma^z\ele{\lambda}{\sigma^z}{\lambda},
\label{acca}
\end{equation}
where $\ket{\lambda}$ is the single-molecule state to be determined
self-consistently by solving the nonlinear eigenvalue problem
associated to (\ref{acca}).
The parameter $G$ represents the dipole interaction energy of a 
single molecule with the rest of the gas.
This must be identified with a sum over all possible molecular 
distances and all possible dipole orientations calculated with the 
Boltzmann factor at temperature $T$.
If $\varrho$ is the density of the gas, we have
\begin{equation}
G=\int_d^\infty  
\frac{\mu^4}{3(4\pi \varepsilon_0\varepsilon_r)^2 k_BT r^6}
~\varrho~4\pi r^2 \mathrm{d}r ,
\label{GG}
\end{equation}
where $\varepsilon_r$ is the relative dielectric constant,  
$d$ the molecular collision diameter and 
the fraction in the integrand represents the Keesom energy between 
two classical dipoles of moment $\mu$ at distance $r$ \cite{Keesom}.
Equation (\ref{GG}) is valid in the high temperature limit which
is appropriate for room temperature experiments.
The exact expression, we assume $\varrho$ constant, is
\begin{eqnarray}
G &=& \frac{\varrho \mu^2}{4\pi \varepsilon_0\varepsilon_r}
\int_{\sqrt[3]{T/T_0}}^\infty
\mathrm{d}x 
~4 \pi x^2
\ln\left[\int_0^1\!\!\mathrm{d}y\int_0^1\!\!\mathrm{d}z \right.
\nonumber \\&&\left.
\times \cosh\left( 2yzx^{-3} \right) 
I_0\left(\sqrt{1-y^2}\sqrt{1-z^2} x^{-3}\right) 
\right],~~~
\label{Gexact}
\end{eqnarray}
where $T_0= \mu^2/(4\pi \varepsilon_0\varepsilon_r d^3 k_B)$
and $I_n(u)$ is the modified Bessel function of the first kind.
Expression (\ref{Gexact}) can be evaluated numerically and 
the result differs very little from (\ref{GG}) even for $T \ll T_0$.
For densities not too high, we set $\varrho=P/k_BT$ so that,
at fixed temperature, the mean-field interaction constant $G$ 
increases linearly with the gas pressure $P$.  
By evaluating (\ref{GG}) we have
\begin{equation}
G=\frac{4 \pi}{9} \left( \frac{T_0}{T}\right)^2 P d^3.
\label{G}
\end{equation}

The solution of the eigenvalue problem associated to the 
Hamiltonian (\ref{acca}) gives the following results.
If $G< \Delta E/2$, there is only one ground state
$\ket{\lambda_0}=\ket{1}$, 
with energy
\begin{equation}
E_0=- \frac{\Delta E}{2}.
\end{equation}
If $ G \geq \Delta E/ 2 $, there are two degenerate ground states
\ba
\ket{\lambda_0^L}&=&\sqrt{\frac{1}{2} +\frac{\Delta E}{4G}}~\ket{1}
+\sqrt{\frac{1}{2}-\frac{\Delta E}{4G}}~\ket{2} 
\label{chiral1}\\
\ket{\lambda_0^R}&=&\sigma^x\ket{\lambda_0^L},
\label{chiral2}
\ea
with energy
\begin{equation}
E_0^L=E_0^R=-\frac{\Delta E}{2}-\frac{1}{2G}
\left( \frac{\Delta E}{2}-G \right)^2.
\end{equation}

By defining the critical value $G_\mathrm{cr}=\Delta E/2$, 
we distinguish the following two cases.
For $G<G_\mathrm{cr}$, 
the ground state of the system is approximated by a 
product of delocalized symmetric single-molecule states 
corresponding to the ground state of an isolated molecule.
For $G \geq G_\mathrm{cr}$, we have two different product states 
which approximate the ground state of the system. 
The corresponding single-molecule states transform 
one into the other under the action of the inversion operator $\sigma^x$, 
see Eq. (\ref{chiral2}), and, for $G \gg G_\mathrm{cr}$, 
they become localized
\begin{equation}
\lim_{\Delta E/G \rightarrow 0}\ket{\lambda_0^L}=\ket{L}
\qquad
\lim_{\Delta E/G \rightarrow 0}\ket{\lambda_0^R}=\ket{R}.
\label{loc}
\end{equation}

The above results imply a bifurcation of the ground state 
at a critical interaction $G=G_\mathrm{cr}$. 
According to Eq. (\ref{G}) and using $\varrho=P/k_BT$, 
this transition can be obtained
by increasing the gas pressure above the critical value 
\begin{equation}
P_\mathrm{cr}=\frac{9}{8 \pi} P_0 \left(\frac{T}{T_0}\right)^2,
\label{pcr}
\end{equation}
where $P_0=\Delta E/d^3$.

When the gas is exposed to an electro-magnetic radiation of
angular frequency $\omega_0$, we add to the Hamiltonian 
(\ref{acca}) the perturbation
\begin{equation}
h_\mathrm{em}(t)=\epsilon f(t) \sigma^z
\label{Hem}
\end{equation}
where $\epsilon$ is a small parameter and 
$f(t)=\theta(t) \cos(\omega_0 t)$,
$\theta(t)$ being the Heaviside function.
The choice of a dipole coupling approximation,
$h_\mathrm{em} \propto \sigma^z$, is justified 
for a radiation of wavelength long with respect to the
molecular size.
Under the effect of the perturbation (\ref{Hem}) 
the single-molecule state $\ket{\lambda(t)}$ evolves according to
the time-dependent nonlinear Schr\"odinger equation
\begin{equation}
i\hbar\frac{\mathrm{d}\ket{\lambda(t)}}{\mathrm{d}t} =
\left[ h(\lambda(t)) + \epsilon f(t) \sigma^z \right]
\ket{\lambda(t)},
\label{hmft}
\end{equation}
with $h(\lambda)$ given by (\ref{acca}).
The linear response to the perturbation (\ref{Hem}) 
is expressed by the generalized susceptibility \cite{Blaizot}
${\mathcal R}(\omega) = \tilde{{\mathcal S}}_1(\omega) /
\tilde{f}(\omega)$,
where $\tilde{f}(\omega)$ and $\tilde{{\mathcal S}}_1(\omega)$ 
are the Fourier transforms of $f(t)$ and ${\mathcal S}_1(t)$,
with ${\mathcal S}_1(t)$ defined by 
\begin{eqnarray}
{\mathcal S} (t) &\equiv&
\ele{\lambda(t)}{\sigma^z}{\lambda (t)}
\nonumber \\
&=& {\mathcal S}_0 (t)+ \epsilon~ {\mathcal S}_1 (t) + \ldots.
\end{eqnarray}

Let us assume that at time $t=0$ each molecule is in the delocalized 
ground state  $\ket{\lambda_0} = \ket{1}$.
The solution of Eq. (\ref{hmft}) with the initial condition
$\ket{\lambda (0)}=\ket{1}$, gives
\begin{equation}
\label{R}
\mathcal{R}(\omega)=
\frac{2 \Delta E}
{(\hbar\omega)^2-
\left( \Delta E^2-2G\Delta E \right)}.
\end{equation}
The generalized susceptibility has a unique pole at positive frequency 
which corresponds to the inversion line frequency 
\begin{equation}
\bar{\nu}=
\frac{\Delta E}{h}\left( 1-\frac{2G}{\Delta E}\right)^{\frac12}.
\label{freq}
\end{equation}
The residue of ${\mathcal R}(\omega)$ at this pole, 
namely $ \left( 1-2G/\Delta E\right)^{-1/2}$,
represents the corresponding transition probability.

\begin{figure}   
\psfrag{P}[][][0.9]{$P~(\mbox{atm})$}
\psfrag{hnu}[B][][0.9]{$h \bar\nu~(\mbox{cm}^{-1})$}
\psfrag{NH_3}[][][0.9]{$NH_3$}
\psfrag{ND_3}[][][0.9]{$ND_3$}
\includegraphics[width=8cm]{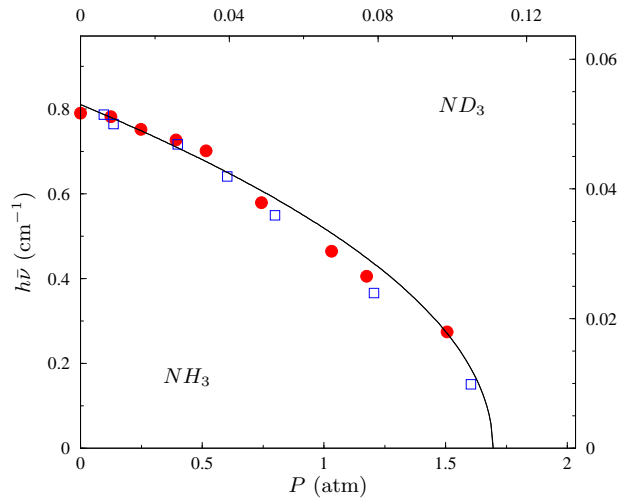}
\\
\caption{Measured inversion-line frequency $\bar\nu$ 
as a function of the gas pressure $P$
for $NH_3$ (dots, left and bottom scales, data from
\protect{\cite{BL1,BL2}})
and $ND_3$ (squares, right and top scales, data from 
\protect{\cite{Birnbaum}}). 
The solid line is the theoretical formula (\protect{\ref{cicacica}}) 
with $P_\mathrm{cr} = 1.695$ atm for $NH_3$ 
and $P_\mathrm{cr} = 0.111$ atm for $ND_3$
calculated according to (\protect{\ref{pcr}}).}
\label{fig1}
\end{figure}

Now we compare our theoretical analysis of the inversion 
line with the spectroscopic data available for 
ammonia \cite{BL1,BL2} and deuterated ammonia \cite{Birnbaum}.
In these experiments the absorption coefficient of a cell
containing $NH_3$ or $ND_3$ gas at room temperature   
was measured at different pressures.
The resulting data are reported in Fig. \ref{fig1}.
The frequency $\bar{\nu}$ of the inversion line decreases by increasing 
$P$ and vanishes for pressures greater than a critical value.
There is factor about 15 between the critical pressures of $NH_3$ and
$ND_3$.
 
By using Eq. (\ref{G}) with $\varrho=P/k_BT$, the theoretical 
expression (\ref{freq}) for the inversion line frequency becomes
\begin{equation}
\bar\nu=\frac{\Delta E}{h}\sqrt{1-\frac{P}{P_\mathrm{cr}}},
\label{cicacica}
\end{equation}
where $P_\mathrm{cr}$ is given by (\ref{pcr}).
Note that this expression does not contain free parameters.
We used the following values taken from \cite{Townes,HCP}: 
$\mu=1.47$ D, $d=4.32$ \AA, $\Delta E_{NH_3}=0.81$ cm$^{-1}$,
$\Delta E_{ND_3}=0.053$ cm$^{-1}$.
Assuming $\varepsilon_r=1$ and $T=300$ K, we obtain
$P_\mathrm{cr} = 1.695$ atm for $NH_3$ 
and $P_\mathrm{cr} = 0.111$ atm for $ND_3$.
The agreement of the theoretical $\bar{\nu}(P)$, also shown in Fig. 1, 
with the experimental data is impressive considering the simplicity
of the model.

Equation (\ref{cicacica}) predicts that, 
up to a  pressure rescaling, the same behavior of $\bar\nu(P)$ 
is obtained for different pyramidal molecules
\begin{equation}
\frac{\bar\nu_{XY_3}(P)}{\bar\nu_{XY_3}(0)}=
\frac{\bar\nu_{X'Y'_3}(\gamma P)}{\bar\nu_{X'Y'_3}(0)},
\label{pgamma}
\end{equation}
where
$\gamma = \left. P_\mathrm{cr}\right._{X'Y'_3}/
\left.P_\mathrm{cr}\right._{XY_3}$.
In the case of $ND_3$ and $NH_3$, at the same temperature $T$ we have 
$\gamma= \Delta E_{NH_3}/\Delta E_{ND_3} \simeq 15.28$.
This factor has been used to fix the scales of Fig. 1.
We see that in this way the $NH_3$ and $ND_3$ data fall on the same curve.

The intensity $I$ of the inversion line predicted by our theoretical 
analysis is given, up to a constant, by the product of the 
photon energy $h \bar{\nu}$ and the residue of (\ref{R}).
The divergence of the transition probability is cancelled by 
the vanishing photon energy and we obtain 
for a gas of $N$ molecules $I \propto N \Delta E$.
By writing $N=PV/k_BT$, 
$V$ being the volume of the absorption cell containing the gas, 
we have that at fixed temperature the line intensity 
increases linearly with pressure. 
This behavior is confirmed by the experimental data 
\cite{BL1,BL2,Birnbaum}.

It is interesting at this point to compare our approach with a
previous study \cite{br} where the experimental data are successfully
reproduced by a formula with three free parameters 
for the shape factor of the inversion line.
In \cite{br} these parameters are determined by trial and error. 
In our work we propose a simplified theory for the shift
of the inversion line based on the dipole-dipole interaction 
and there are no free parameters.
The specific prediction of our model for the critical pressure
$P_\mathrm{cr}$ in terms of the electric dipole $\mu$ of the molecule, 
its size $d$, the splitting $\Delta E$ and the temperature $T$ 
of the gas, successfully verified in the case of ammonia,  
should be experimentally tested also for other pyramidal gases.
We emphasize that the study of the inversion spectra of pyramidal 
molecules, like $NH_3$ and $ND_3$, 
in recent years has acquired a considerable interest in 
geophysical \cite{spilker} and astrophysical research \cite{henkel}.

It is not easy to assess the region of validity of the model 
starting from first principles, i.e. considering all the degrees of
freedom of the molecules.
It is reasonable to assume that the mean field calculation is 
meaningful as long as the interactions among the molecules are 
small compared to the inversion line frequency (\ref{freq}).
If we take the model seriously, at least for the qualitative aspects, 
at pressures greater than the critical one we have the following 
situation.
In the limit of an infinite number of molecules, 
the Hilbert space separates into two sectors generated by the ground 
state vectors given in mean field approximation by
\ba
\ket{\psi_0^L}&=&\ket{\lambda_0^L}\ldots\ket{\lambda_0^L}\\
\ket{\psi_0^R}&=&\ket{\lambda_0^R}\ldots\ket{\lambda_0^R},
\ea
These sectors, which we call $\mathcal{H}_L$ and $\mathcal{H}_R$,  
cannot be connected by any operator involving a finite number of 
degrees of freedom (local operator). 
According to \cite{Wightman} this means that a superselection rule
operates between the two sectors distinguished by the eigenvalues 
of an observable.
It is natural \cite{jpt} to define the chirality operator 
\begin{equation}
\chi=\lim_{N \to \infty} \frac{1}{N}\sum_{i=1}^N 
\mathbbm{1}_1\otimes\ldots \otimes\sigma^z_i 
\otimes\ldots\otimes\mathbbm{1}_N.
\end{equation}
It is immediately verified that 
\begin{equation}
\langle\psi|\chi|\psi\rangle = \pm~
\sqrt{1-\left(\frac{\Delta E}{2G}\right)^2}
\end{equation}
for $\psi$ in $\mathcal{H}_L$ and $\mathcal{H}_R$, respectively.
This mean value distinguishes the two sectors but is a property 
weaker than $\psi$ being an eigenstate of $\chi$.
According to (\ref{loc}) only in the limit $\Delta E/G \to 0$, 
the states $\psi$ become
completely localized and therefore eigenstates of $\chi$.
The fact that the $\psi$ are not eigenstates of $\chi$ 
for $\Delta E/G \neq 0$
is connected to the non-orthogonality of the mean-field 
one-molecule states $\ket{\lambda_0^L}$ and $\ket{\lambda_0^R}$
and is presumably an artifact of the approximation.

This analysis has an interesting implication.
Our model applies not only to molecules $XY_3$ but also
to their substituted derivatives $XYWZ$. 
In fact, the difference should not matter as far as the shift 
of the inversion line is concerned.  
However, an important difference between the two cases is that for 
$XY_3$ the localized states can be obtained one from the other either 
by rotation or by space inversion, while for $XYWZ$ they can be 
connected only by space inversion. 
This implies that $XYWZ$ molecules at a pressure greater 
than the critical value are chiral and therefore optically active.
Recent ultrasensitive experimental methods have been developed 
which allow the quantitative measurement of optical rotation 
in gaseous compounds \cite{vaccaro}. 
It would be interesting to measure the optical activity 
of pyramidal gases for $P>P_\mathrm{cr}$.

\begin{acknowledgments}
One of us (G. J.-L.) acknowledges a great debt to a previous 
collaboration with Pierre Claverie.
We thank A.S. Wightman for emphasizing on many occasions the conceptual
importance of the problem discussed and
C. Di Castro, F. Gianturco and J. Lebowitz for stimulating discussions.
This work was supported in part by Cofinanziamento MURST protocollo 
MM02263577\_001.
\end{acknowledgments}

\end{document}